\documentclass[apj]{emulateapj}
\journalinfo{{\rm Journal reference:
  {\sc The Astrophysical Journal,} 635:1126--1135, 2005 December 20}}
\slugcomment{Received by ApJ 2004 May 28, accepted 2005 July 30,
published 2005 December 20}

\usepackage{mathptmx}

\newcommand{\vB}{{\bf B}}

\newcommand{\lfrac}[2]{{{#1}/{#2}}}
\newcommand{\rhomean}{{\bar{\rho}}}
\newcommand{\EK}{{E_K}}
\newcommand{\seed}{{\rm Seed}}
\newcommand{\LAD}{{L_{\rm{AD}}}}
\newcommand{\tni}{{t_{ni}}}
\newcommand{\nion}{{n_i}}
\newcommand{\cs}{{c_s}}
\newcommand{\ts}{{t_s}}
\newcommand{\tg}{{t_g}}
\newcommand{\Ls}{{L_s}}
\newcommand{\nHtwo}{{n_{{\rm H}_2}}}
\newcommand{\vA}{{v_{\rm{A}}}}
\newcommand{\rhoS}{{\rho_{\rm{S}}}}
\newcommand{\phiS}{{\phi_{\rm{S}}}}
\newcommand{\MS}{{M_{\rm{S}}}}
\newcommand{\nJ}{{n_{\rm{J}}}}
\newcommand{\LJ}{{L_{\rm{J}}}}
\newcommand{\MJ}{{M_{\rm{J}}}}
\newcommand{\NT}{{N_{\rm{T}}}}
\newcommand{\rhoT}{{\rho_{\rm{T}}}}
\newcommand{\tT}{{t_{\rm{T}}}}
\newcommand{\tten}{{t_{10}}}

\newcommand{\cm}{{\rm\,cm}}

\newcommand{\pc}{{\rm\,pc}}

\newcommand{\Myr}{{\rm\,Myr}}

\newcommand{\kms}{{\rm\,km\,s^{-1}}}

\newcommand{\msun}{{\rm\,M_\odot}}

\newcommand{\K}{{\rm\,K}}
\newcommand{\second}{{\rm\,s}}

\begin{document}

\title{Nonlinear Criterion for the Stability of Molecular Clouds}

\author{Ruben Krasnopolsky and Charles F. Gammie}

\affil{Center for Theoretical Astrophysics,
University of Illinois at Urbana-Champaign, Loomis Laboratory of Physics,\\
1110 West Green Street, Urbana, IL 61801}

\shortauthors{{\sc Krasnopolsky and Gammie}}
\shorttitle{{\sc Nonlinear Stability of Molecular Clouds}}

\begin{abstract}

Dynamically significant magnetic fields are routinely observed in
molecular clouds, with mass-to-flux ratio $\lambda \equiv (2\pi\sqrt{G})
\Sigma/B \sim 1$ (here $\Sigma$ is the total column density and $B$ is
the field strength).  It is widely believed that ``subcritical'' clouds with
$\lambda < 1$ cannot collapse, based on virial arguments by
Mestel and Spitzer and a linear
stability analysis by Nakano and Nakamura.  Here we confirm, using high
resolution numerical models that begin with a strongly supersonic
velocity dispersion, that this criterion is a fully nonlinear stability
condition.  All the high-resolution models with $\lambda \le 0.95$ form
``Spitzer sheets'' but
collapse no further. All models with $\lambda \ge 1.02$ collapse to the
maximum numerically resolvable density.  We also investigate other
factors determining the collapse time for supercritical models.  We show
that there is a strong stochastic element in the collapse time: models
that differ only in details of their initial conditions can have
collapse times that vary by as much as a factor of 3.
The collapse time cannot be determined from just the velocity dispersion;
it depends also on its distribution.  Finally, we discuss the
astrophysical implications of our results.

\end{abstract}


\section{Introduction}

Molecular clouds evolve under the influence of self-gravity so as to
condense part of their mass into dense cores and, ultimately, stars.
The presence of magnetic fields can prevent or delay condensation.  The
possibility was first studied by \cite{ms56}, who noted that the magnetic
energy and the gravitational energy scale in exactly the same way with
the radius $R$ of the cloud ($\propto 1/R$) if flux freezing obtains.
They argued that there was therefore a critical mass below which a cloud
threaded by a particular field strength would be unable to collapse.

A more precise but less general argument was advanced by \cite{nn78},
who studied the linear theory of a self-gravitating, isothermal,
equilibrium sheet of plasma threaded by a perpendicular magnetic field.
They found that magnetic fields stabilize the sheet against
gravitational collapse if the mass-to-flux ratio is smaller than
$1/2\pi\sqrt{G}$.  

These results motivate the definition of a dimensionless mass-to-flux
ratio, 
\begin{equation}
\lambda \equiv 2\pi\sqrt{G} {\Sigma\over{B}},
\end{equation}
where $\Sigma$ is the column density of the sheet, and $B$ is the
magnetic field strength.  The exact coefficient used to define $\lambda$
depends somewhat on the geometry of the collapse.  Here we have chosen
the coefficient most relevant to the magnetic field geometry adopted in
this paper, tending to produce thin sheets, in agreement with the
expectations for magnetically supported clouds.  Clouds with $\lambda >
1$ are termed {\it supercritical}, and clouds with $\lambda < 1$ are
termed {\it subcritical}.

Both the \citeauthor{ms56} and the \citeauthor{nn78} models consider
exact equilibria.  Molecular clouds are far from equilibrium, however,
with near-virial, highly supersonic velocity dispersion.  These internal
velocities must arise from strong turbulence.\footnote{The most
plausible alternative to turbulence, some type of weakly dissipative
ordered flow, does not emerge naturally in any relevant numerical
experiments that we are aware of.  The mode-mode coupling is always
strong.  Even circularly polarized Alfv\'en waves, which are exact
solutions to the compressible equations of motion, suffer from a
parametric instability with a dynamical decay rate \citep{sag69,gol78}.}
Turbulence might change the stability properties of the cloud, either by
compressing a $\lambda < 1$ flow until it collapses, or by providing
turbulent support to a cloud with $\lambda > 1$.

Many works have suggested that turbulence could provide support to
star-forming clouds.  \citet{cf53a} included turbulent support in their
model for interstellar gaseous structures.  \citet{ms56} pointed out
that turbulence tends to decay, and that turbulence of amplitude large
enough to support a cloud against self-gravity would decay especially
quickly, although allowing the possibility that a strong magnetic field
might perhaps allow longer lived turbulence.  The supersonic linewidths
observed in molecular clouds were attributed to radial motions inside
the cloud instead of turbulence by \cite{gk74}.  \citet{zp74} argued
that if this interpretation were true for all clouds where such
fluctuations are observed, the star formation rate would be too large by
at least one order of magnitude.  \citet{am75} then suggested that the
observed velocity fluctuations are due to hydromagnetic waves.  By the
late 1980s, this idea was widely accepted \citep[e.g.,][]{sal87}.  In the
late 1990s, however, a succession of numerical experiments
\citep{ml98,sog98,go96} strongly suggested that the damping time of
turbulence in
magnetized molecular clouds is close to the dynamical time.  If one
accepts this, then turbulent pressure can be effective in supporting
self-gravitating clouds only if it is constantly replenished, in which
case the support is perhaps more readily identified with the
stirring mechanism than with the turbulence itself.

Other work has tended to emphasize the role of turbulence in initiating
gravitational collapse \citep[e.g.,][]{mk04}.  Regions with a convergent
velocity field will naturally tend to collapse sooner than regions with
divergent velocity fields.  It seems highly likely that some parts of
molecular clouds have strongly convergent velocity fields; is this ever
enough to overcome the stabilizing effects of the magnetic field?
Can a subcritical cloud be induced to collapse by squeezing, or can a
supercritical cloud be prevented from collapsing by the introduction of
turbulence?  The purpose of this paper is to investigate these questions
using a simple series of numerical experiments.

The plan of the paper is as follows.  In \S 2 we describe the
experimental design, our numerical methods, and the diffusion
characteristics of our code (based on the ZEUS algorithm).
In \S 3 we describe results, including a ``fiducial'' run,
and the influence of physical and numerical parameters on the outcome.
\S 4 summarizes and discusses astrophysical implications.

\section{Description of numerical experiments}
\label{description}

We will consider the simplest possible system that can manifest
sub/supercritical behavior: a two-dimensional, periodic box containing a
magnetized, self-gravitating, isothermal gas.  Since we are interested in
studying the effects of turbulence, we will introduce a velocity field
in the initial conditions with statistical properties similar to those
found in interstellar clouds.  We will then allow the system to evolve
for many dynamical times, or until it ``collapses.'' 

Specifically, we consider a square domain in the $x-y$ plane of size $L
\times L$.  The $z$ direction points out of this plane; no quantity depends
on $z$.  The initial fluid density is $\rhomean$, and the sound speed,
which is constant in space and time, is $\cs$.  The initial field is
${\bf B} = B_x {\hat {\bf x}}$, where $B_x$ is constant.  The strength
of the field can be characterized by $\lambda = 2 \pi \sqrt{G}
\rhomean L/B_x$.  We set the initial value of $\left< B_z \right>=0$,
because otherwise in this $z$-independent geometry,
asymptotically there is no collapse.

The initial velocity field is a Gaussian random field with zero
divergence, constructed as in \cite{ogs99}.  The initial velocity field
has a power spectrum $\left<v_k^2\right>\propto k^{-3}$ for
$2\pi/L<k<8\,(2\pi/L)$ and ${\bf k} \cdot {\bf v}_k = 0$.  This power
spectrum is consistent, in 2D, with Larson's Law $v_\lambda \sim
\lambda^{1/2}$, which is equivalent to an energy spectrum $E_k \sim
k^{-2}$ (in 3D Larson's Law
implies $\left<v_k^2\right>\propto k^{-4}$).
The velocity is normalized so that the kinetic energy $\EK$ matches the
desired value, and $1/3$ of the kinetic energy is in motions
perpendicular to the $x - y$ plane of the simulation.

\subsection{Spitzer sheets and simulation units}

In our experiments we will frequently find that matter flows along
magnetic field lines to form sheets normal to the field.  These sheets
are given coherence by the self-gravity of the medium.  \cite{spitz42}
was the first to consider the problem of the vertical structure of an
infinite, self-gravitating, isothermal sheet.  Spitzer's solution turns
out to be highly useful in understanding the evolution of our
simulations.  Spitzer found that the equilibrium density profile of a
sheet of surface density $\Sigma$ is $\rho(z) = (\Sigma/2 H) {\rm
sech}^2(z/H)$ where $H = \cs^2/(\pi G \Sigma)$.  The corresponding
gravitational potential and field are $\phiS = 2 \cs^2
\log(\cosh(z/H))$ and $g = -2\pi G \Sigma \tanh(z/H)$.

For any given Spitzer sheet formed during our simulation the surface
density parameter $\Sigma$ corresponds to $\rhomean \Ls$, where $\Ls$ is
the extent of the region along the fieldlines that a given sheet has
collected its mass from.  At late times we can assume that $\Ls\approx
L$: most of the mass originally distributed along the fieldlines will be
collected into the given sheets.  This is true for the most massive
sheets in the supercritical simulations, and it is seen even more
clearly in the subcritical simulations, where at late times in the
simulation a single large scale, stable sheet incorporates most of the
mass of the system.  In the following we will assume $\Ls\approx L$ and
$\Sigma=L\rhomean$ for the Spitzer sheets we are largely interested in
--- those that have collected most mass.

We nondimensionalize our models by setting $L = 1, \rhomean = 1$, and
$\cs = 1$.  The simulation time unit is therefore $L/\cs$, the sound
crossing time.  In these units, Newton's gravitational constant equals
$\pi\nJ^2$, the Jeans length is $1/\nJ$, the peak density $\rhoS$ of
an equilibrium Spitzer sheet of $\Ls=L$ is $(\pi\nJ)^2/2$, and its
half-thickness $H$ equals $(\pi\nJ)^{-2}$.  Notice that, because of the
periodic boundary conditions the Spitzer sheets are slightly distorted,
but as long as $H \ll 1$, the Spitzer solution will be approximately
correct.

Most of the simulations presented in this paper have $\nJ=3$.  This implies
that the semithickness of the Spitzer sheet is $0.011$, and if we are to
resolve this with at least four grid zones we need $N > 360$, where the
resolution of our uniform grid is $N^2$.  This rather stringent
resolution requirement explains why we have chosen to study the problem
in 2D rather than 3D.

This Spitzer-sheet model is especially useful when $\lambda\lesssim 1$
inside a largely ordered magnetic field, able to channel the flow into sheets,
which later might become unstable and collapse, through accretion, collision,
and merger.
This sheet model, however, is not useful where $\lambda\gg 1$ and
collapse is unconstrained by the field.

Simulation units can be converted to dimensional values by assuming for
illustrative purposes a typical density $\nHtwo=10^2\cm^{-3}$,
and a typical temperature $T=10\K$.  Then the sound speed $\cs=0.19\kms$
fixes the unit of speed, and from the Jeans length
$\LJ=\cs(\pi/G\rhomean)^{1/2}=1.9\pc$ we obtain the unit of length
$L=\nJ\LJ=5.7\pc$ for our standard value $\nJ=3$; a Spitzer sheet would
then have a peak density of $\nHtwo=4.4\times10^3\cm^{-3}$, and
$H=0.064\pc$.  The unit of mass is given by $\rhomean L^3=\rhomean
\nJ^3\LJ^3 =1.3\times 10^3\msun$.  The unit of time is the sound
crossing time $\ts=L/\cs \approx 30 \Myr$.  A characteristic
gravitational contraction time is $\tg=\LJ/\cs\approx 10\Myr$; free-fall
collapse times are on the order of $0.3 \tg\approx 3\Myr$, depending on
the geometry of the collapse.

The sound-crossing time of a Spitzer sheet is $\sim H/\cs\approx
0.3\Myr$.  From the Spitzer sheet parameters $\rhoS$ and $H$ it is
possible to define a characteristic ``Spitzer'' mass $\MS = (\pi H)^2
\Sigma = \cs^4/(G^2 \Sigma) = (\pi^2\nJ^4)^{-1}$ in dimensionless
units, with $\Sigma= L\rhomean$.  The factor of $\pi^2$ is designed to
capture the mass inside half a wavelength of the shortest unstable mode
of the sheet \citep{led51,ee78}.  For parameters typical of a molecular
cloud, $\MS = 1.6\msun$, which is a suggestive result.  This may be
compared with the thermal Jeans mass $\MJ = \rhomean \LJ^3
=\lfrac{\pi^{3/2} \cs^3}{G^{3/2} \rhomean^{1/2}} =\nJ^{-3}$ in
dimensionless units, with a typical value of $\MJ= 49\left(\lfrac{T}{10
\K}\right)^{3/2} \left(\lfrac{\nHtwo}{10 \cm^{-3}}\right)^{-1/2}
\msun$.

\subsection{Ambipolar diffusion}\label{AD}
Our simulation utilizes the ideal MHD equations, which have some
well-known limitations.
Ambipolar diffusion is expected to become relevant
\citep{kp69} at lengthscales
smaller than the damping length for Alfv\'en waves,
$\LAD\sim \vA \tni$, where $\vA=B/\sqrt{4\pi\rho}$,
and $\tni=1/(K \nion)$, with $K\approx 1.9\times 10^{-9}\cm^3\second^{-1}$
\citep{drd83}.
The number density of ions $\nion$ may depend strongly on environmental
factors, such as the UV illumination and its attenuation by
the cloud material \citep{cm95,mk89}; it also depends on chemical properties,
such as the metal abundance in the gas phase.
For our fiducial mean density $\rhomean$, a representative
value could be $\nion \sim 2\times 10^{-4}\cm^{-3}$, largely limited
by an assumed metal abundance $x_M\approx 10^{-6}$.
For the typical peak density of a Spitzer sheet $\rhoS$ in our
conditions, $\nion \sim 6\times 10^{-4}\cm^{-3}$
for cosmic-ray dominated ionization, and about ten times larger for regions
of the cloud that are moderately well UV-illuminated.

The ambipolar diffusion lengthscale $\LAD$ can now be compared with
the size $L$ of the computational volume, giving
an estimate of the scale at which ideal MHD stops
being a complete dynamical description of the flow.
For our mean density profile, we find that this lengthscale is
$\LAD\sim L/35$, much larger than our typical grid spacing $L/512$,
and comparable to our typical sheet thickness $2H=L/45$.
However, ideal MHD is still a good description of the most important portions
of this study; the regions where mass is collected to form dense sheets.
For $\rho=\rhoS$, the lengthscales are $\LAD\sim L/700$ for a UV-dark region,
and $\LAD\sim L/7000$ for the more illuminated case.  The decrease of
$\vA$ with density has contributed to this effect, together with the
larger $\nion$.

We keep in mind, however, that densities and ionization rates vary
widely inside clouds, and so the ambipolar diffusion lengthscales
and timescales may vary widely.
Turbulent conditions inside the flow are also expected to increase the
importance of ambipolar diffusion, even at larger lengthscales, especially
near sharp velocity and magnetic gradients.

\subsection{Stopping criterion}

We must fix some criterion for stopping the numerical integration; when
the density in any zone equals or exceeds the largest allowed by the
Truelove numerical stability condition \citep{truelove97} the run is
terminated and classified as having collapsed.  The Truelove condition
requires that the local Jeans length be resolved by some
algorithm-dependent number $\NT$ of grid
zones, typically about $4$.  This requirement sets a maximum resolvable
density of $\rhoT=(N/\nJ\NT)^2 = 1820 (N/512)^2$ for $\nJ=3$
on our uniform
grid of $N^2$ zones.  We have found that further integration of
Truelove-unstable models results in large local fluctuations in the
density, which can produce ``explosions'' that corrupt the entire
computational domain.  Runs that reach $t = 2$ without violating the
Truelove condition are classified as stable.

We have experimented with other collapse detection schemes, because the
Truelove criterion has the deficiency that it is resolution dependent.
In the Tables described below we report not only the time $\tT$ at which
the Truelove condition is violated, but also the time $\tten$ when
$1\%$ of the mass exceeds $10$ times the Spitzer density $\rhoS$.
These times are typically close to each other, and both can be
considered as measures of the onset of gravitational instability.  The
time $\tten$ has the advantage of not depending explicitly on numerical
resolution, but it can be fooled into producing misleadingly short
collapse times by strong density fluctuations, particularly when the
turbulent kinetic energy is large.

\subsection{Numerical methods and tests}

Our simulations are run on a fixed 2D Cartesian grid, using the ZEUS
algorithm \citep{sn92a,sn92b} as implemented for instance in
\citet{ogs99}.  ZEUS is a numerical algorithm to evolve ideal
(non-resistive, non-viscous) non-relativistic MHD flows.  It is
operator-split, representing the fields on a (possibly moving) Eulerian
staggered mesh.  The magnetic field evolution uses constrained transport
\citep{eh} which guarantees that $\nabla\cdot\vB=0$ to machine
precision, combined with the method of characteristics \citep{hs}, which
ensures accurate propagation of Alfv\'en waves.  ZEUS is explicit in
time, and so the timestep $\Delta t$ is limited by the Courant
conditions.  In our problem, usually the most stringent has been
$\Delta t < \Delta x/\vA$, where $\vA=B/\sqrt{4\pi\rho}$
can take very large values in density-depleted regions.
A numerical density floor, $\rho_{\rm floor}$, has been set to limit
density depletion, preventing $\Delta t$ from becoming too small;
we have directly tested that this tiny non-conservation of mass
by the code does not alter the simulation results regarding collapse
in any way.
The Poisson equation, needed to describe self-gravity, is solved by
Fourier transform methods, using the FFTW code \citep{fftw}.

Any Eulerian scheme will cause some diffusion of the magnetic field with
respect to the mass.  It is crucial for our experiment that this
nonconservation of $\lambda$ be as small as possible.  The numerical
diffusivity of ZEUS is difficult to estimate because, unlike a physical
resistivity, it is flow dependent.  An empirical approach is therefore 
required.  

We have studied conservation of $\lambda$ using two distinct methods.
In the first method we initialize a non--self-gravitating box using the
same initial data as in our main experiments, as described above.  We
evolve the computation to $t = 0.5$ and then damp the velocity field
exponentially (with timescale $t_{\rm damp} = 0.05$) until $t = 10$.  If
there were no $\lambda$ diffusion the box would return to a uniform
density, uniform field state.  Diffusion changes $\lambda$, so the final
state consists of a unidirectional magnetic field with density and field
strength varying only perpendicular to the field, from which $\lambda$
can be easily measured.

In the second method we initialize a non--self-gravitating box using the
same initial data as in our main experiments, but we evolve the
computation only to $t = 0.5$.  We then sample 80 field lines chosen to
lie at equal intervals of the vertical component of the vector potential
(equivalent to lines equally spaced in magnetic flux).  We then
integrate $\rho/\left(B_x^2 + B_y^2\right)^{1/2}$
along the field line in the $x - y$ plane,\footnote{This is equivalent
to integrating $\rho/\left(B_x^2 + B_y^2 + B_z^2\right)^{1/2}$ along
the 3D fieldline.}
using linear interpolation to determine $\rho$ and ${\bf B}$ at each
position, which immediately yields $\lambda$.

These two methods give nearly identical results.  We therefore adopt the
second method exclusively, since it can be used to probe existing
numerical data without any additional, expensive evolution.

One possible figure of merit for the diffusion in $\lambda$ is
$\sigma_\lambda/\lambda_0$, where $\sigma_\lambda$ is the dispersion in
sampled values of $\lambda$ at the final instant of the simulation,
and $\lambda_0$ is the nominal initial value (which we call simply
$\lambda$ outside of this subsection).
Tables \ref{table:lambdavsN} and \ref{table:lambdavst}
show $\sigma_\lambda/\lambda_0$ as a function of
resolution and of time during a single simulation, respectively.  The
run shown in Table~\ref{table:lambdavst}
has a resolution of $512^2$.  The key points here
are that $\sigma_\lambda/\lambda_0$ decreases as resolution increases,
and that in every case $\sigma_\lambda/\lambda_0$ is about $10\%$ or
less, which suggests that we should be able to measure the critical
value of $\lambda$ to similar accuracy.  

Evidently $\sigma_\lambda/\lambda_0$ is converging, but as $\approx
N^{-1/2}$ rather than the expected $N^{-1}$.  This may be because of the
existence of unresolved regions in the flow where most of the diffusion
occurs, or it may be the result of irreducible ``turbulent'' diffusion
that is present independent of the magnitude of the effective numerical
diffusion.
The numerical diffusion is correlated with the amplitude of turbulence.
According to Table~\ref{table:lambdavst}, much of the diffusion occurs
very early in the run, when the rms velocity is large.

\begin{deluxetable}{rllll}[b]
\tablecaption{
Mass-to-flux diffusion in ZEUS,
as a function of numerical resolution
\label{table:lambdavsN}}
\tablehead{
\colhead{$N$} &
\colhead{$t$} &
\colhead{$\sigma_\lambda/\lambda_0$} &
\colhead{$\lambda_{\rm max}/\lambda_0$} &
\colhead{$\lambda_{\rm min}/\lambda_0$} }
\startdata
400  & 0.1    & 9.0 \% & 1.26 & 0.78  \\ %
512  & 0.1    & 8.6 \% & 1.31 & 0.78  \\ %
1024 & 0.1    & 6.1 \% & 1.30 & 0.87  \\ %
2048 & 0.1    & 4.4 \% & 1.10 & 0.86  \\ %
\tableline
512  & 0.2    & 8.3 \% & 1.20 & 0.79  \\ %
2048 & 0.2    & 4.7 \% & 1.12 & 0.85     %
\enddata
\end{deluxetable}

\begin{deluxetable}{lrllr}
\tablecaption{
Mass-to-flux diffusion in ZEUS, as a function of time
\label{table:lambdavst}}
\tablehead{
\colhead{$t$} &
\colhead{$\sigma_\lambda/\lambda_0$} &
\colhead{$\lambda_{\rm max}/\lambda_0$} &
\colhead{$\lambda_{\rm min}/\lambda_0$} &
\colhead{$\EK(t)$}}
\startdata
0     &  0\phantom{.0}\% 
               & 1       & 1      & 50   \\ %
0.01  &  0.2\% & 1.01    & 0.99   & 37   \\ %
0.02  &  0.9\% & 1.02    & 0.97   & 25   \\ %
0.03  &  2.1\% & 1.05    & 0.92   & 18   \\ %
0.04  &  3.6\% & 1.10    & 0.90   & 15   \\ %
0.05  &  6.4\% & 1.18    & 0.84   & 16   \\ %
0.06  &  8.0\% & 1.32    & 0.74   & 19   \\ %
0.1   &  8.6\% & 1.31    & 0.78   & 17   \\ %
0.2   &  8.3\% & 1.20    & 0.79   & 11   \\ %
0.3   &  9.6\% & 1.20    & 0.74   & 8.5  \\ %
0.34  & 11.1\% & 1.21    & 0.71   & 5.5     %
\enddata
\end{deluxetable}

Another possible figure of merit is the total variation in $\lambda$,
that is, $|\lambda_{\rm max} - \lambda_{\rm min}|/(2 \lambda_0)$,
measuring the possible existence of localized diffusion events in
addition to the overall diffusivity of the code.
We find in Tables~\ref{table:lambdavsN} and \ref{table:lambdavst}
that this quantity is typically of the size $\sim 2.4 \sigma_\lambda$,
expected in the mean for the half-range of a sample of 80 elements randomly
taken from a Gaussian distribution.
However, the distribution of values of $\lambda$ might not always be Gaussian,
because localized numerical diffusion could be important for some fieldlines.
In our tables, this may be happening when
$\lambda_{\rm max}$ reaches values as large as $1.3\lambda_0$,
even at the relatively high resolution of $N=1024$.
The larger total variation of $\lambda$ observed in those
simulations suggests a possible risk of masking the
subcritical nature of some models.\footnote{As a possible example of
this effect, in Table~\ref{table:resolution}
one model collapses despite having $\lambda_0=0.9$;
also the very low resolution
models with $N=128$ that collapse for $\lambda_0=0.744$.}
However, we also find that the total variation of $\lambda$
starts to drop at the even higher resolution of $N=2048$;
numerical resolution seems apparently able to reduce also this
more local measure of the diffusivity of the numerical code.

It is worth noting that ambipolar diffusion in nature is likely strong
enough to dominate the diffusion measured here.  For our nominal cloud
parameters, we have seen in \S 2.2 that the damping length of Alfv\'en
waves is of the same order as the expected thickness of an equilibrium sheet.
Thus our models may misrepresent the situation in nature by tying the
fluid too closely to the magnetic field.  The combination of
ambipolar diffusion and turbulence (which can drive sharp features for
the ambipolar diffusion to act on) may be a potent driver of variations
in mass-to-flux ratio in Galactic molecular clouds.

\section{Results}
\label{results}

\subsection{Fiducial run} 

As a guide to the dynamics of our numerical experiments, we will first
describe a ``fiducial'' run, whose behavior is in a sense typical of the
other experiments.  This run is supercritical, with $\lambda=1.5$,
initial $\EK=50$ (equivalent to an rms Mach number of $10$), $\nJ=3$,
and $N=512$.  The panels of Figures \ref{fig:fiducial}
and \ref{fig:fiducial_lines} show how condensation
proceeds.  At first, small density concentrations form due to both ram
pressure associated with the supersonic velocity fluctuations in the
initial conditions and fluctuations in the magnetic pressure.  These
later coalesce into larger clumps, typically oriented perpendicular to
the magnetic field.\footnote{Clumps do not tend to orient
perpendicular to the field in our three dimensional models \citep{glso03}.
Those runs had a resolution of $256^3$, however, and the
supercritical runs did not have $\lambda$ as close to $1$ as the models
considered here.} We have seen that these clumps develop into fully
stable Spitzer sheets in the simulations of subcritical clouds; here
they can be considered also as approximate Spitzer sheets, which later
come unstable, as the peak density of these sheets grows.  This density
growth takes place when the clumps merge or collide, and when matter
accretes from outside the sheet.
The largest density of these clumps increases as shown in
Fig.~\ref{fig:fiducial_max_density}; slowly at the beginning,
but very steeply close to the end of the run.
The energies, on the other hand, vary smoothly in time
(Fig.~\ref{fig:fiducial_energies}), and are not a good predictor
of the time required for instability.

At a time $\tten=0.325$, the fraction of matter denser than 10 times
the nominal Spitzer density includes more than 1\% of the mass
$\rho_{1\%}(t)>10\rhoS$.  Not long afterwards, at a time $\tT=0.341$
(soon after the last panel in Fig.~\ref{fig:fiducial}) the peak density
of the simulation box exceeds the Truelove limit, forcing an end to the
run.  We conclude that the initial state of this fiducial run represents
an unstable cloud, able to produce dense cores.  These two times are
much larger than the linear e-folding time found in \citet{nakano88}
($\approx 0.035$ for $\lambda=1.5$); collecting matter into the unstable
structures takes a longer time than the instability process, linear or
nonlinear, and dominates the total time necessary to achieve
instability in the mildly supercritical clouds.  Using our nominal
conversion factors from simulation to physical units, $\tten = 9.8\Myr$.

\subsection{The nonlinear stability criterion}
\label{stabilitycriterion}

To discover how precisely the criticality condition was obeyed in the
numerical experiments, we considered a series of runs with $\EK = 50$
and $\EK = 10$ while gradually varying $\lambda$.
Table~\ref{table:stability} lists the
collapse times for each of these runs.  Evidently the criticality
condition is very nearly obeyed in the numerical evolutions, and there
is no evidence of collapse induced by compression.  Indeed, given the
diffusion of $\lambda$ measured in \S 2,
it is remarkable (from a
numerical standpoint) that we are able to reproduce the condition so
accurately.  Some sense of the ``error bars'' can be obtained by
noticing that the $\lambda = 1$ model with $\EK = 50$ does collapse,
while the $\lambda = 1.02$ model with $\EK = 10$ does not.  This
suggests that the Nakano \& Nakamura condition is the true nonlinear
stability condition.

Tables~\ref{table:turbulence} and \ref{table:resolution} show a clear
trend to make the simulations shorter lived as $\lambda$ increases.
This trend is expected from the already observed
stability criterion.\footnote{The most strongly supercritical
models ($\lambda\gtrsim 10$) collapse very quickly, in around one
free-fall time.}  Models with $\lambda \approx 1$ can be quite long
lived; the $\lambda = 1$, $\EK = 50$ model persists until $t = 0.756$,
or about $23\Myr$ for our nominal cloud parameters.

The run with $\lambda=1.05$ and $\EK=10$ has been done twice, with
different values of the numerical density floor; the collapse times
are identical up to reasonable precision.  This allows us to trust the
runs using the larger density floor, which are much more convenient
because it allows a larger timestep, largely
controlled by the maximum value of the Alfv\'en speed
$B/\sqrt{4\pi\rho}$ on the grid.  From here on, we will not report the
values of this purely numerical parameter in these simulations.

\subsection{Influence of the turbulence energy and distribution}

We have also investigated the effect of the amplitude and structure of
the initial velocity field.  Table~\ref{table:turbulence}
shows the results from a series of runs with
$\nJ = 3$ and $N = 512$.
The column marked ``Seed'' is the seed used to
initiate the random number generator used to generate the initial
velocity field.  Runs with the same seed but
different initial kinetic energies have velocity fields that are
linearly proportional to each other.  

The first series of runs with $\rm{seed}=1$ show a monotonic increase in the
lifetime of the cloud with kinetic energy, consistent with results
reported elsewhere \citep{go96, ogs99}.
The effect is weak at low energies but more pronounced once $\EK > 50$.  

An even larger effect is obtained by changing the structure of the
initial velocity field, i.e.\ by changing the seed.  We find that models
that differ only in the initial seed can have collapse times that vary
by up to a factor of $3$.

To further explore this effect, we performed a series of runs, choosing
sixty different values of the random seed used to set up the shape of
the initial velocity distribution.  In this series, we have fixed
$\lambda=1.5$, $\EK=50$, $N=512$, and $\nJ$=3.  The results
can be seen in Figure~\ref{fig:distribution}, showing a wide distribution
of collapse times.  The total range of this sample goes from $\tT=0.228$ to
$\tT=0.667$, equal to $\sim 7$ to $20\Myr$ for the typical cloud
parameters used in \S 2.  The mean time is $\langle \tT
\rangle=0.368$ ($\sim 11\Myr$); the median is located at $\tT=0.355$,
and the peak near $\tT=0.3$, showing a moderate asymmetry.  This
asymmetry is more pronounced in the tails: $\tT<0.2$ is not observed in
the sample; while a few values of $\tT>0.5$ (at a similar distance from
the median but in the opposite direction) are present in the
distribution, corresponding to clouds lasting between $\sim 15$ and
$20\Myr$ before collapse, much longer than the mean lifetime value.

This stochastic variation in cloud lifetime doubtless has a counterpart
in nature.  The origin of this variability is clear: almost all velocity
variations occur at the largest scales, and are driven by just a few
Fourier modes.  If these modes happen to have the right amplitude and
phase then collapse is hastened.  If they are unfavorable, then collapse
can be delayed by as much as $12\Myr$ for our nominal cloud parameters.

\begin{deluxetable}{lllll}
\tablecaption{
Models with $\lambda$ close to 1
\label{table:stability}}
\tablehead{
\colhead{$\lambda$} & \colhead{$\EK$} &
\colhead{$\rho_{\rm floor}$} &
\colhead{$\tten$}  &  \colhead{$\tT$} } 
\startdata
1.1     &  50 &  $10^{-6}$      & 0.304   & 0.376  \\  
1.05    &  50 &  $10^{-4}$      & 0.387   & 0.500  \\  
1.0     &  50 &  $10^{-4}$      & 0.497   & 0.756  \\  
0.95    &  50 &  $10^{-4}$      & $>2$    & $>2$   \\  
0.9     &  50 &  $10^{-4}$      & $>2$    & $>2$   \\  
\tableline
1.1     &  10 &  $10^{-6}$      & 0.314   & 0.367  \\  
1.05    &  10 &  $10^{-6}$      & 0.702   & 0.741  \\  
1.05    &  10 &  $10^{-4}$      & 0.702   & 0.741  \\  
1.02    &  10 &  $10^{-4}$      & $>2$    & $>2$   \\  
1.0     &  10 &  $10^{-4}$      & $>2$    & $>2$   \\  
1.0     &  10 &  $10^{-6}$      & $>2$    & $>2$   \\  
0.9     &  10 &  $10^{-6}$      & $>2$    & $>2$       
\enddata
\tablecomments{
Parameters kept fixed in these runs: $\nJ=3$, $N=512$,
random $\rm{seed} = 2$.}
\end{deluxetable}

\begin{deluxetable}{lrcll}
\tablecaption{
Supercritical and subcritical models
\label{table:turbulence}}
\tablehead{
\colhead{$\lambda$} & \colhead{$\EK$} &
\colhead{$\seed$} &
\colhead{$\tten$}  &  \colhead{$\tT$} } 
\startdata
1.5     & 100 & 1    & 0.601   & 0.615 \\  
1.5     & 70  & 1    & 0.521   & 0.533 \\  
1.5     & 50  & 1    & 0.325   & 0.341 \\  
1.5     & 20  & 1    & 0.260   & 0.270 \\  
1.5     & 10  & 1    & 0.250   & 0.257 \\  
1.5     & 1   & 1    & 0.259   & 0.269 \\  
\cline{2-4}
1.5     & 50  & 2    & 0.126   & 0.211 \\  
1.5     & 20  & 2    & 0.193   & 0.200 \\  
1.5     & 10  & 2    & 0.233   & 0.244 \\  
\cline{2-4}
1.5     & 50  & 3    & 0.510   & 0.537 \\  
1.5     & 20  & 3    & 0.346   & 0.360 \\  
1.5     & 10  & 3    & 0.299   & 0.310 \\  
1.5     & 1   & 3    & 0.314   & 0.324 \\  
\tableline
1.2     & 100 & 1    & 0.715   & 0.744 \\  
1.2     & 50  & 1    & 0.376   & 0.400 \\  
\tableline
1.1     & 100 & 1    & 0.787   & 0.910 \\  
1.1     & 50  & 1    & 0.377   & 0.534 \\  
\cline{2-4}
1.1     & 100 & 2    & 0.314   & 0.372 \\  
1.1     & 50  & 2    & 0.304   & 0.376 \\  
1.1     & 10 &  2    & 0.314   & 0.367 \\  
\cline{2-4}
1.1     & 50  & 3    & 0.735   & 0.746 \\  
\tableline
0.8     & 100 & 2    & 0.216   &  $>2$  \\ 
0.8     & 10  & 2    & $>2$    &  $>2$  \\ 
\cline{2-4}
0.8     & 10  & 3    & 0.664   &  $>2$  \\ 
0.8     & 1   & 3    & $>2$    &  $>2$     
\enddata
\tablecomments{
Parameters kept fixed in these runs: $\nJ=3$, $N=512$.
}
\end{deluxetable}

\begin{deluxetable}{lrcrll}
\tablecaption{Supercritical and subcritical models
run at different numerical resolutions
\label{table:resolution}}
\tablehead{
\colhead{$\lambda$} & \colhead{$\EK$} &
\colhead{$\seed$} & \colhead{$N$}    &
\colhead{$\tten$}  &  \colhead{$\tT$} } 
\startdata
$\infty$ & 50 & 1    & 512  & 0.068   & 0.077  \\ 
$\infty$ & 50 & 1    & 1024 & 0.071   & 0.083  \\ 
\tableline
1000    & 50  & 1    & 512  & 0.068   & 0.077  \\ 
\tableline
10      & 50  & 1    & 512  & 0.112   & 0.124  \\ 
\tableline
1.5     & 50  & 1    & 256  & 0.283   & 0.283  \\ 
1.5     & 50  & 1    & 400  & 0.305   & 0.312  \\ 
1.5     & 50  & 1    & 512  & 0.325   & 0.341  \\ 
1.5     & 50  & 1    & 1024 & 0.409   & 0.434  \\ 
1.5     & 50  & 1    & 1536 & 0.429   & 0.453  \\ 
1.5     & 50  & 1    & 2048 & 0.445   & 0.481  \\ 
\cline{2-5}
1.5     & 20  & 1    & 512  & 0.260   & 0.270 \\  
1.5     & 20  & 1    & 1024 & 0.263   & 0.283  \\ 
\cline{2-5}
1.5     & 10  & 1    & 512  & 0.250   & 0.257  \\ 
1.5     & 10  & 1    & 1024 & 0.253   & 0.273  \\ 
\tableline
1.2     & 50  & 1    & 256  & 0.319   & 0.319  \\ 
1.2     & 50  & 1    & 512  & 0.376   & 0.400  \\ 
1.2     & 50  & 1    & 1024 & 0.504   & 0.523  \\ 
\tableline
1.1     & 100 & 2    & 512  & 0.314   & 0.372  \\ 
1.1     & 100 & 2    & 1024 & 0.337   & 0.738  \\ 
\tableline
0.9     & 100 & 2    & 512  & 0.079   & 0.351  \\ 
0.9     & 100 & 2    & 576  & 0.079   & $>2$   \\ 
\cline{2-5}
0.9     & 100 & 1    & 512  & $>2$    & $>2$   \\ 
\tableline
0.85    & 100 & 2    & 512  & 0.075   & $>2$   \\ 
\tableline
0.8     & 100 & 2    & 512  & 0.216   & $>2$   \\ 
\cline{2-5}
0.8     &  50 & 1    & 256  & $>2$    & $>2$   \\ 
\tableline
0.7     & 100 & 2    & 512  & 0.435   & $>2$   \\ 
\tableline
0.6     &  50 & 1    & 256  & $>2$    & $>2$      
\enddata
\tablecomments{
Parameter kept fixed in these runs: $\nJ=3$.
}
\end{deluxetable}

\begin{deluxetable}{lrcrll}
\tablecaption{Models with different values of $\nJ$.
\label{table:nj}}
\tablehead{
\colhead{$\nJ$} & \colhead{$\lambda$} &
\colhead{$\seed$} & \colhead{$N$}    &  \colhead{$\tT$} } 
\startdata
2.5     & 1.2 & 1    & 512  & 0.87  \\ 
        & 1.2 & 2    & 256  & 0.44  \\ 
        & 1.2 & 3    & 256  & 0.83  \\ 
\cline{2-5}
        & 1.1 & 1    & 256  & 0.75  \\ 
        & 1.1 & 2    & 256  & 0.48  \\ 
        & 1.1 & 3    & 256  & 0.82  \\ 
        & 1.1 & 3    & 512  & 1.21  \\ 
        & 1.1 & 4    & 256  & 0.49  \\ 
\cline{2-5}
        & 0.9 & 3    & 512  & $>2$  \\ 
\tableline
2.0     & 1.2 & 1    & 256  & 1.77  \\ 
        & 1.2 & 1    & 512  & 1.87  \\ 
        & 1.2 & 2    & 256  & 1.00  \\ 
        & 1.2 & 3    & 256  & 1.50  \\ 
\cline{2-5}
        & 1.15& 2    & 256  & 1.33  \\ 
        & 1.15& 3    & 256  & $>2$  \\ 
\cline{2-5}
        & 1.1 & 1    & 256  & $>2$  \\ 
        & 1.1 & 1    & 512  & $>2$  \\ 
        & 1.1 & 2    & 256  & 9.77  \\ 
        & 1.1 & 2    & 512  & $>2$  \\ 
        & 1.1 & 3    & 256  & $>2$  \\ 
        & 1.1 & 3    & 512  & $>2$  \\ 
        & 1.1 & 4    & 256  & $>2$  \\ 
\cline{2-5}
        & 0.9 & 3    & 512  & $>2$     
\enddata
\tablecomments{
Parameter kept fixed in these runs: $\EK=50$.
}
\end{deluxetable}

\subsection{Influence of numerical parameters}

The influence of the density floor $\rho_{\rm floor}$ has already been
shown in \S 3.2 to be fully negligible, provided
this floor is not unreasonably large.

Numerical resolution, on the other hand, can be quite relevant.  Any
serious simulation of condensation in a nearly critical cloud
must be able to resolve the
half-thickness $H$ of a Spitzer sheet, requiring at the very minimum
$N>1/H =(\pi\nJ)^2 $.  For our fiducial choice of $\nJ$=3, this requires a
minimum of $N > 89$, and more reasonably $N>200$; any simulation run at
smaller resolution would not be exploring the most basic physics of mass
condensation.  However, this requirement does not seem to
take care of all the effects of numerical resolution.

Table~\ref{table:resolution} lists simulations where we have
varied the number $N$ of active zones in the grid on each direction.
There is a clear tendency for the more resolved simulations to delay
$\tT$.  This was in part expected, as the Truelove limit density
$\rhoT=(N/\nJ\NT)^2$ depends steeply on $N$.  This is not, however,
the main reason for the observed trend.  Peak densities grow very quickly
in the neighborhood of the condensation time, almost nullifying in most
cases the influence of the exact magnitude of $\rhoT$ on the value of
$\tT$.  The quantity $\tten$ is expected to be less directly
dependent on resolution, having a definition where $N$ does not appear;
it still shows some dependence on resolution, closely correlated to the
dependence shown by $\tT$, and probably due to details of the dynamics
being more revealed at higher resolutions and to reduced numerical
diffusion.

Most of our simulations were performed at $N=512$.
For comparison purposes, a simulation run with the same initial
conditions as the fiducial run, but with $N=2048$, has $\tT=0.445$ and
$\tten=0.481$, noticeably larger, but not enough to change the
qualitative conclusions.

One of the subcritical models with $\lambda=0.9$ and $N=512$
(Table~\ref{table:resolution})
is remarkable because the run presents an unexpected collapse
at $\tT=0.351$.
However, the same table shows that either changing the numerical seed,
or a moderate increase in numerical resolution is enough to suppress
this unusual behavior; also a small change in $\lambda$
can suppress this apparently purely artificial collapse.

We have also run a few comparison simulations at $N=128$.  This resolution
is insufficient to represent equilibrium Spitzer sheets,
and indeed the models collapse even for $\lambda=0.744$.  This is due
to the excessively low resolution: it is just enough to accommodate one sheet
semithickness per grid zone, and, through the
Truelove stability criterion, it allows only a narrow density range,
limited by $\rhoT=(N/\nJ\NT)^2=114\rhomean=2.6\rhoS$,
insufficient to accommodate
an eventual moderately large oscillation in the density $\rhoS$
of the equilibrium sheets formed in subcritical simulations.

In another series of tests, we started the simulations from an
equilibrium sheet, and let it evolve in the presence
of a very small perturbation.  For the values $\lambda=1.5$
and $\lambda=2$, the perturbation grows
linearly with $e$-folding times equal to $0.036\pm 0.001$ and
$0.027\pm 0.001$, quite comparable to the values predicted
by the linear theory, $0.035$ and $0.025$, known from
\citet{nakano88} to a precision of $5$\%.
For values of $\lambda<1$, the simulations have remained stable
up to a time $t\approx 4$ during a few linear test runs
performed at $\lambda=0.9$ at various resolutions.

\subsection{Influence of $\nJ$}
We have performed a small set of simulations exploring the
influence of $\nJ$ on the instability criterion found (Table~\ref{table:nj}).
For $\nJ=3$, we had found instability for $\lambda>1.05$;
for $\nJ=2.5$, the numerical requirement is no more stringent than
$\lambda>1.1$;
however, for $\nJ=2.0$, the numerical requirement for instability
becomes $\lambda>1.15$.
A weak dependence of the numerical criterion on $\nJ$ had already
been predicted by \citet{nakano88} in the linear regime.  Linearly
unstable modes have a minimum critical wavelength; if we require
that this wavelength must fit inside the computational box size $L$,
we find that the instability criterion will be approximately\footnote{
Following \citet{nakano88}, certain
integrals involving generalized Riemann zeta functions have been
replaced by simpler expressions.
These approximations are excellent
inside our range of interest $\nJ\geq 2$.}
\begin{equation}\label{eq:nj}
\lambda> \left[1 - 2/(\pi\nJ^2)\right]^{-1}\ ,
\end{equation}
which for small values of $\nJ$ can be more stringent than the
infinite disk value $\lambda>1$.
The results in Table~\ref{table:nj} are consistent with Eq.~\ref{eq:nj}.
\citet{nakano88} presents for the case of a finite disk
a still more stringent criterion for linear instability,
$\lambda> \left[1 - 4/(\pi\nJ^2)\right]^{-1}$,
based on the assumption that {\em two}
critical wavelengths should fit inside the computational box.
Our slightly larger unstable range might be related to the geometric
difference between a finite disk and our periodic boundary conditions.
Spitzer sheets pull in magnetic field lines
during their formation and contraction (Fig.~\ref{fig:fiducial_lines});
this may also allow collapse at smaller wavelengths than expected in a
purely linear theory.

\section{Conclusions}

Our simulations confirm that the single most important element in
determining the long term gravitational stability of turbulent
magnetized clouds is indeed the mass-to-flux ratio, dividing
supercritical from subcritical clouds.
The relevant coefficient is that corresponding to a sheet geometry,
as derived by \citet{nn78}.

Turbulent energy has comparatively little influence on the presence or
absence of stability, up to Mach numbers $\sim 10$.
Subcritical clouds will develop density concentrations due to
this turbulence, but under an ideal MHD regime, the consequent
increase in magnetic pressure prevents further collapse.
However, total turbulent energy has some influence on the lifetime of
supercritical clouds, especially as the Mach number becomes large
enough (of the order of $\sim 7$ in these simulations).

More interesting is the fact that turbulence introduces a stochastic
element.  The collapse time cannot be predicted with certainty from
physical parameters such as the mass and field in the cloud, and the
typical energy of the turbulence motions, because the random
distributions of velocity and density can change the lifetime by some
factor, seen to be of the order of 3 in one large sample.
The resulting distribution of lifetimes has an asymmetric tail of
unusually long-lived clouds.  We suggest that the existence of such a
tail may introduce a bias in the observed samples of star-forming clouds.
Most star formation will take place in the more frequent,
shorter lived clouds, while observations of clouds will tend to focus
on the fewer longer lived ones.

We have seen that the numerical resolution requirements needed to
study cloud collapse are very stringent, and we expect they will be even
more stringent in 3D.
There is a necessity of resolving the possible equilibrium structures,
such as the Spitzer sheets, which we have seen fully formed in the
subcritical clouds, and partially formed during the run-up to
instability of the mildly supercritical ones.
The thickness of these sheets scale with the number $\nJ$ of
Jeans lengths as $\nJ^{-2}$.
Accommodating a large number $\nJ$ of Jeans lengths
inside the computational volume will therefore be numerically challenging.
Increasing $\nJ$ by only a factor of 2 requires increasing the space
resolution by a factor of 4.
Unless adaptive mesh refinement (AMR) is used, this requires
increasing the simulation runtime by factors on the order of $64=4^3$
in 2D, and $256=4^4$ in 3D.
We anticipate that AMR will be used in many of the
successful simulations of core formation in the future.

Numerical stability, through the Truelove condition, sets a maximum
density that can be accommodated at a given spatial resolution.
Shocks in strongly turbulent flows have large compression ratios,
sometimes requiring increasing resolution in order to distinguish a
transient density increase due to a shock from an authentically
unstable accumulation of mass able to form a collapsed object.

We have seen that artificially enforcing numerical density floors,
even relatively large ones, on the order of $10^{-4}$ times the
background density, had almost no influence in the evolution of the
collapse.
This result is again not surprising, because wide regions of small
density have little influence on the dense, self-gravitating regions
that undergo collapse.  Density floors can significantly speed up ideal MHD
simulations, whose Courant timestep is often limited by large
Alfv\'en speeds $B/\sqrt{4\pi\rho}$ in the least dense regions.

This work is limited due to the periodic boundary conditions.
We believe this may have favored the collection of clumps
into larger clumps until the instability can take
place.  Some simulations occasionally show fast-moving clumps
flowing past each other, and later merging
once one of them returns through the other side of the periodic
computational volume.  The periodic boundary conditions
make it plausible that sooner or later, most of the mass in a given
fieldline will collect into a single clump, which then
can undergo instability if its mass is even slightly supercritical.
In real clouds with ordered magnetic fields, clumps inside the same
fieldline but moving in opposite directions are not expected to merge;
however, it is improbable this will apply to all of the fieldlines and
so we expect that the instability will still take place in a similar form,
albeit with an additional stochastic factor in the cloud lifetime.

Two-dimensionality is also a limitation of this work.
It has strongly limited the topological possibilities for the fieldlines;
it is conceivable that the consequent limitations in motion have favored
the collection of mass into massive sheets and other structures.
Observations \citep[e.g.,][]{gbmm90, crutcher04}, and
3D simulations and studies \citep[e.g.,][]{basu00, glso03}
indeed indicate that sheets aligned perpendicular to the magnetic field
are not always the preferred possibility for the long term development
of clouds.  More variety of clump shapes is expected in a 3D study.
The larger variety in motions allowed by a 3D magnetic field
is expected to enhance the already observed stochastic effects, and
perhaps might also delay mass collection into potentially unstable structures.
However, even in 3D, the simulations performed by \citet{osg01}
suggest that the stability criterion will still be dominated by the
mass-to-flux ratio.

In some of our models, artificial numerical diffusion has
turned an initially uniform mass-to-flux ratio $\lambda$
into a non-uniform distribution, sometimes with striking
effects on the numerical stability.
While this has a numerical origin, non-uniform
distributions of mass-to-flux are also expected on astrophysical grounds.
For instance, turbulence provides structures and shocks
with small lengthscales and strong magnetic gradients,
conditions favorable to a localized, efficient ambipolar diffusion,
which can redistribute mass and magnetic flux independently.
Cloud collisions can also merge together portions of gas
having different masses and magnetic fields.
We plan to study directly the physical effect of a
non-uniform mass-to-flux ratio in our future work.

\acknowledgments

This work was supported by NASA grant NAG 5-9180.  We thank Jon McKinney,
Eve Ostriker, Zhi-Yun Li, and Chris Matzner for comments.

{\vskip 15ex
\plotone{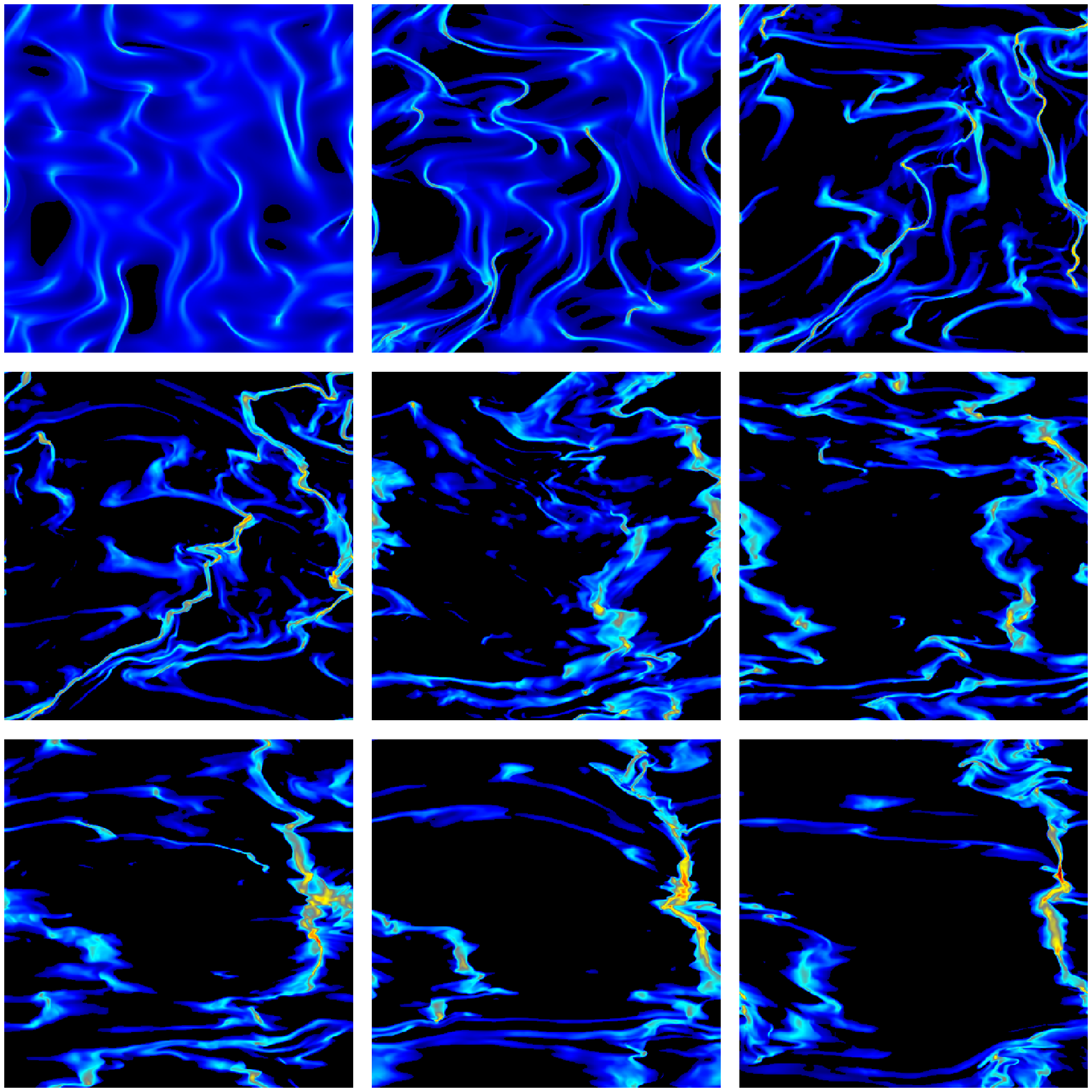}}
{\figcaption[f1.eps]{\label{fig:fiducial}
Colormaps of $\rho$ in the fiducial run.  Snapshots at times $t=0.01$,
$0.02$, $0.04$, $0.06$, $0.12$, $0.18$, $0.24$, $0.30$, and $0.34$.
The logarithmic
colorscale goes from dark blue (saturating on black) to deep red,
corresponding to densities going from $0.01\rhoS = 0.4441\rhomean$ to
$10\rhoS=444.1\rhomean$, where $\rhoS$ is the peak density of an
equilibrium Spitzer sheet for the given parameters.
}}

\plotone{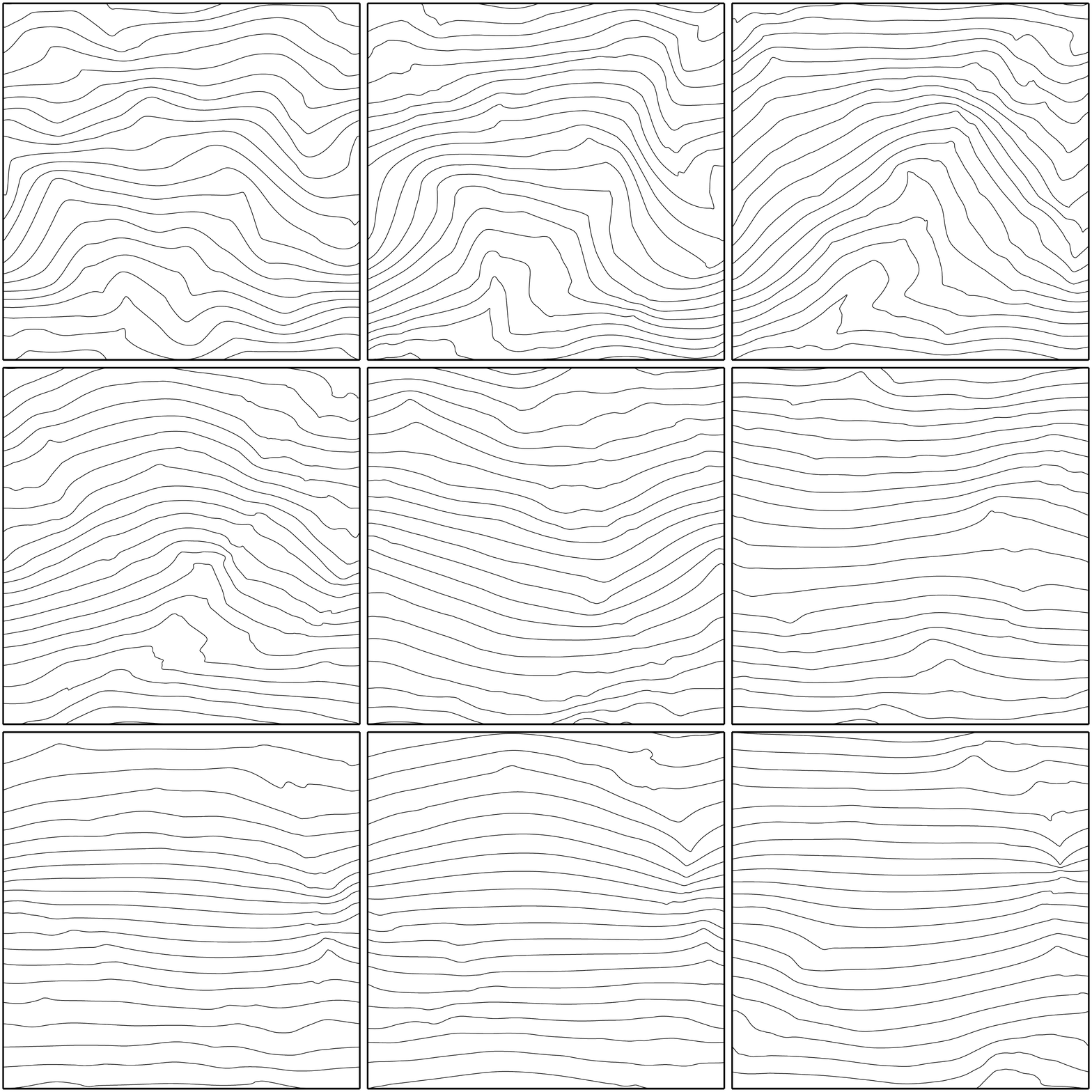}
{\figcaption[f2.eps]{\label{fig:fiducial_lines}
Field lines in the fiducial run.  The figure shows snapshots at times $t
= 0.01$, $0.02$, $0.04$, $0.06$, $0.12$, $0.18$, $0.24$, $0.30$, and $0.34$.
}}

\plotone{f3.eps}
{\figcaption[f3.eps]{\label{fig:fiducial_max_density}
Increase of the maximum mass density $\rho_{\max}$ with time
in the fiducial run.  The run finishes when $\rho_{\max}=1820$,
the Truelove value, at a time $t=0.341$, slightly beyond the plotted
region.
Some of the transient peaks shown here could have provoked a numerical
instability at a resolution smaller than the value $N=512$
adopted for the fiducial run.
}}

\plotone{f4.eps}
{\figcaption{\label{fig:fiducial_energies}
Variations in time of the turbulent kinetic energy $\EK$, the total
magnetic energy $E_B$, and (minus) the gravitational energy
$-E_G$.
}}

{\quad\\ \vskip 15ex
\plotone{f5.eps}}
{\figcaption{\label{fig:distribution}
Frequency of the different values of $\tT$.  Sixty simulations have been
run with the parameters $\lambda=1.5$, $\EK=50$, $N=512$, and different
random seeds.  Each of these runs has reported a value of $\tT$.  This
plot was constructed by summing 60 Gaussian profiles (with
$\sigma=0.03 \ts=0.9\Myr$) centered at each of these $\tT$ values.
Collapse times range from $\sim 7$ to $20\Myr$.
}}


\begin{thebibliography}{}

\bibitem[Arons \& Max(1975)]{am75}
Arons, J., \& Max, C. E. 1975,
\apj, 196, L77

\bibitem[Basu(2000)]{basu00}
Basu, S. 2000,
\apj, 540, L103

\bibitem[Chandrasekhar \& Fermi(1953)]{cf53a}
Chandrasekhar, S., \& Fermi, E. 1953,
\apj, 118, 113

\bibitem[Ciolek \& Mouschovias(1995)]{cm95}
Ciolek, G. E., \& Mouschovias, T. Ch.\ 1995,
\apj, 454, 194

\bibitem[Crutcher(2004)]{crutcher04}
Crutcher, R. M. 2004,
in The Magnetized Interstellar Medium,
eds.\  B. Uyaniker, W. Reich \& R. Wielebinski,
(Copernicus GmbH: Katlenburg-Lindau), 123

\bibitem[Draine, Roberge, \& Dalgarno(1983)]{drd83}
Draine, B. T., Roberge, W. G., \& Dalgarno, A. 1983,
\apj, 264, 485

\bibitem[Elmegreen \& Elmegreen(1978)]{ee78}
Elmegreen, B. G., \& Elmegreen, D. M. 1978,
\apj, 220, 1051 

\bibitem[Evans \& Hawley(1988)]{eh}
Evans, C. R., \& Hawley, J. F. 1988,
\apj, 332, 659

\bibitem[Frigo \& Johnson(2005)]{fftw}
Frigo, M., \& Johnson, S. G. 2005,
Proc.\ IEEE, 93, 216

\bibitem[Gammie et al.(2003)]{glso03}
Gammie, C. F., Lin, Y.-T., Stone, J. M., \& Ostriker, E. C. 2003,
\apj, 592, 203

\bibitem[Gammie \& Ostriker(1996)]{go96}
Gammie, C. F., \& Ostriker, E. C. 1996,
\apj, 466, 814

\bibitem[Goldreich \& Kwan(1974)]{gk74}
Goldreich, P., \& Kwan, J. 1974,
\apj, 189, 441

\bibitem[Goldstein(1978)]{gol78}
Goldstein, M. L. 1978,
\apj, 219, 700

\bibitem[Goodman et al.(1990)]{gbmm90}
Goodman, A. A., Bastien, P., Myers, P. C., \& M{\'e}nard, F. 1990,
\apj, 359, 363

\bibitem[Hawley \& Stone(1995)]{hs}
Hawley, J. F., \& Stone, J. M. 1995,
Comput.\ Phys.\ Commun., 89, 127

\bibitem[Kulsrud \& Pearce(1969)]{kp69}
Kulsrud, R., \& Pearce, W. P. 1969,
\apj, 156, 445

\bibitem[Ledoux(1951)]{led51}
Ledoux, P. 1951,
Ann.\ d'Astrophys.\ 14, 438

\bibitem[Mac Low et al.(1998)]{ml98}
Mac Low, M.-M., Klessen, R. S., Burkert, A., \& Smith, M. D. 1998,
\prl, 80, 2754

\bibitem[Mac Low \& Klessen(2004)]{mk04}
Mac Low, M.-M., \& Klessen, R. S., 2004
Rev.\ Mod.\ Phys., 76, 125

\bibitem[McKee(1989)]{mk89}
McKee, C. F. 1989,
\apj, 345, 782

\bibitem[Mestel \& Spitzer(1956)]{ms56}
Mestel, L., \& Spitzer, L., Jr.\ 1956, 
\mnras, 116, 503

\bibitem[Nakano \& Nakamura(1978)]{nn78}
Nakano, T., \& Nakamura, T. 1978,
\pasj, 30, 671

\bibitem[Nakano(1988)]{nakano88}
Nakano, T. 1988,
\pasj, 40, 593

\bibitem[Ostriker, Gammie, \& Stone(1999)]{ogs99}
Ostriker, E. C., Gammie, C. F., \& Stone, J. M. 1999,
\apj, 513, 259

\bibitem[Ostriker, Stone, \& Gammie(2001)]{osg01}
Ostriker, E. C., Stone, J. M., \& Gammie, C. F. 2001,
\apj, 546, 980

\bibitem[Sagdeev \& Galeev(1969)]{sag69}
Sagdeev, R. Z., \& Galeev, A. A. 1969,
Nonlinear Plasma Theory (New York: W. A. Benjamin)

\bibitem[Shu, Adams, \& Lizano(1987)]{sal87}
Shu, F. H., Adams, F. C., \& Lizano, S.\ 1987,
\araa, 25, 23 

\bibitem[Spitzer(1942)]{spitz42}
Spitzer, L., Jr.\ 1942,
\apj, 95, 329 

\bibitem[Stone \& Norman(1992a)]{sn92a}
Stone, J. M., \& Norman, M. L. 1992,
\apjs, 80, 753

\bibitem[Stone \& Norman(1992b)]{sn92b}
Stone, J. M., \& Norman, M. L. 1992,
\apjs, 80, 791

\bibitem[Stone, Ostriker, \& Gammie(1998)]{sog98}
Stone, J. M., Ostriker, E. C., \& Gammie, C. F. 1998,
\apjl, 508, L99 

\bibitem[Truelove et al.(1997)]{truelove97}
Truelove, J. K., Klein, R. I., McKee, C. F.,
Holliman, J. H., II,
Howell, L. H.,
\&
Greenough, J. A.
1997,
\apj, 489, L179

\bibitem[Zuckerman \& Palmer(1974)]{zp74}
Zuckerman, B., \& Palmer, P. 1974,
\araa, 12, 279

\end{thebibliography}
\end{document}